# Constructing 100 MΩ and 1 GΩ Resistance Standards via Star-Mesh Transformations

Dean G. Jarrett[*], Albert F. Rigosi[*], Dominick S. Scaletta[‡], Ngoc Thanh Mai Tran[*#], Heather M. Hill[*], Alireza R. Panna[*], Cheng Hsueh Yang[*†], Yanfei Yang[*], Randolph E. Elmquist[*], David B. Newell[*]

[*]National Institute of Standards and Technology, 100 Bureau Drive, Stop 8171, Gaithersburg, MD, 20899, USA

dean.jarrett@nist.gov

[‡]Department of Physics, Mount San Jacinto College, Menifee, CA 92584, USA
[#]Joint Quantum Institute, University of Maryland, College Park, MD 20742, USA
[†]Graduate Institute of Applied Physics, National Taiwan University, Taipei 10617, Taiwan

*Abstract* — A recent mathematical framework for optimizing resistor networks to achieve values in the MΩ through GΩ levels was employed for two specific cases. Objectives here include proof of concept and identification of possible apparatus limitations for future experiments involving graphene-based quantum Hall array resistance standards. Using fractal-like, or recursive, features of the framework allows one to calculate and implement network designs with substantially lower-valued resistors. The cases of 100 MΩ and 1 GΩ demonstrate that, theoretically, one would not need more than 100 quantum Hall elements to achieve these high resistances.

## I. Introduction

Quantized Hall array resistance standard (QHARS) devices are a means of greatly expanding access to many quantized values of resistance and have not historically been used heavily in calibrations. Previous research has yielded QHARS based on GaAs/AlGaAs heterostructures [1] and graphene [2], where customized series and parallel configurations of numerous Hall bar elements are selected based on a desired resistance. More recently, QHARS devices valued at nearly 1 MΩ, with all elements quantized at the $i = 2$ plateau, were measured in a wye-delta (Y-Δ) network configuration to yield an *effective* resistance of about 20.6 MΩ [3]. Standardization of these quantities has a broad importance to physics, including measurements of electrical properties like resistivity and Hall effect, and enable precise understanding of a material's electronic structure and behavior [2] – [6].

Though Y-Δ networks offer some fabrication relief for resistances through the 100 MΩ level, a mathematical framework theorized that using generalized star-mesh devices including recursive designs (or multi-stage designs) drastically reduced the number of required graphene-based elements needed for achieving resistances up to 1 EΩ [2].

Before designing and fabricating such devices, it is important to understand that these high resistance measurements can be made reliably with artifact resistors. One can still mathematically conclude that using lower-valued resistors in a recursive star-mesh configuration may yield much higher effective resistances. In this work, two high-valued resistances are explored, each using a different measurement technique.

In the first case, an effective resistance near 100 MΩ is measured via dual source bridge by constructing wire wound resistors that emulate a potential QHARS device containing 20 elements in series in each the high and low branches, along with 19 parallel single-element branches that are grounded. The second case is a recursive design to obtain nearly 1 GΩ, which is measured with a teraohmmeter and, if built into a QHARS device, would only need 37 quantum Hall elements (see Fig. 1) [3].

These measurements seek to validate: (1) the mathematical framework one may use for minimizing the resistor values needed in a recursive star-mesh design to achieve a particular high-valued resistance, and (2) the use of two measurement methods for such high effective resistances in preparation for future QHARS device implementation. Ultimately, the validation of this framework is important for highlighting the untapped potential of more complex resistance networks for use in metrology.

## II. STAR-MESH TRANSFORMATION

Scaletta *et al.* recently adopted a mathematical approach for minimizing the number of required quantum Hall elements to achieve a desired *effective* value, what has also been labeled a virtual resistance $R_{ij}$ (and when divided by the Hall resistance $R_H$ at the $i = 2$ plateau, the *coefficient of effective resistance* $q_{M:ij}$) [3]. The first principle is [4]:

$$R_{ij} = R_i R_j \sum_{\alpha=i}^{N} \frac{1}{R_\alpha}$$

(1)

Figure 1 (a) shows this virtual-to-real conversion, where one may achieve $R_{ij}$ using real components $R_i$, $R_j$, and $R_k$. Since QHR elements must be represented by whole numbers, it was assumed that grounded branches like $R_k$ cannot be smaller than approximately 12906.4037 Ω. Expanding a virtual resistor to its "real" components is an act of recursion that is defined by the integer $M$ [2]. Figure 1 (b) shows what happens when one embeds virtual resistors within virtual resistors to create a more elaborate, yet real, resistor network (where each of the dark blue branches goes to ground).

In addition to knowing $M$, one then defines $\xi$ as the number of distinct grounded branches (where each branch value is not smaller than 1 quantum Hall element - for a Y-Δ network, $\xi = 1$), along with the total number $D_T$ of quantum Hall elements in the QHARS device [2]:

$$D_T(M, \xi, q_{M:ij}) = \frac{2^M}{\xi}(\xi q_{M:ij} + 1)^{2^{-M}} - \frac{2^M}{\xi} + (2^M - 1)\xi$$

(2)

Equation 2, for the case of the 1 GΩ desired value, may be minimized to find the configuration shown in Fig. 1 (e), which is color coded to correspond to Fig. 1 (c). For that arrangement, composed of four sets of 7 elements in series, with each set containing a fork to a grounded branch (of which there are 3, each containing 3 parallel elements), the effective resistance of nearly 1.01 GΩ may theoretically be measured.

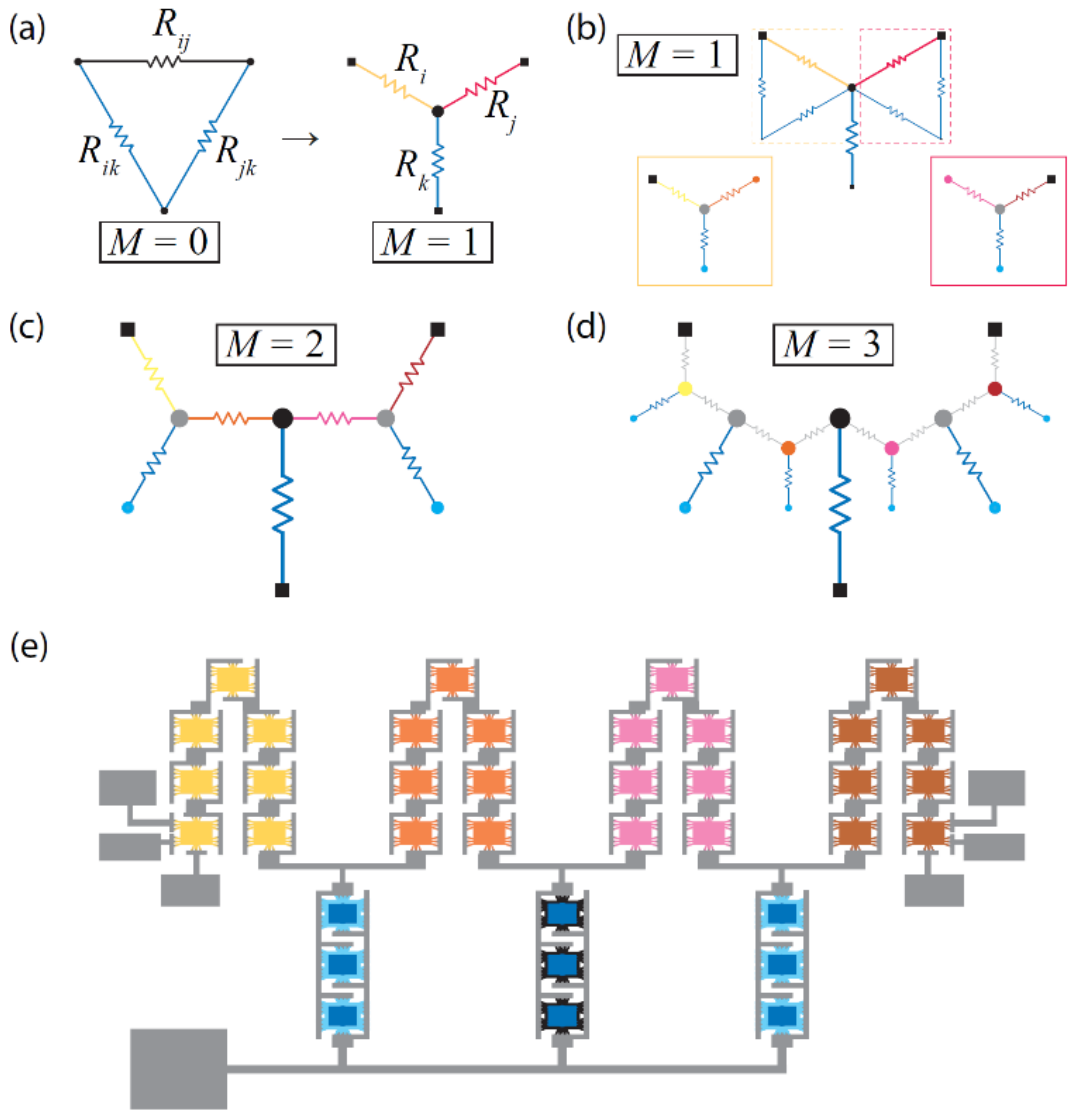

Fig. 1. (a) $R_{ij}$ (or $q_{ij}$) as a Y-Δ network. Every additional expansion of all non-grounded elements increases the characteristic recursion factor $M$ by one. (b) Every resistor in the non-grounded path is expanded as a Y-Δ network, with a second and third iteration shown in (c) and (d), respectively. (e) Optimizing (minimizing) the total number of quantum Hall elements for 1 GΩ yields this device design (with an output of nearly 1.01 GΩ), which is color coded to match (c).

III. RESULTS AND CONCLUSION

Using a dual source bridge and a 100 MΩ Hamon standard, two artifact resistor networks were constructed to emulate 100 MΩ and 1 GΩ. In the first case, $R_i$ and $R_j$ represented 20 quantum Hall elements in series. Using a digital multimeter, $R_i$ and $R_j$ were independently measured to be 7 Ω and 14 Ω larger than $20*R_H$. The grounded branches (19 elements, essentially in parallel) were measured to be about 2 Ω larger than $R_H/19$. Due to the minor changes in each of the branches, the final expected value is 98.358 MΩ. When measured in the star-mesh configuration described in Ref. [3], the effective resistance was approximately 97.7 MΩ (yielding an error of less than 1%). Measurements at 10 V, 20 V, and 50 V are shown in Fig. 2.

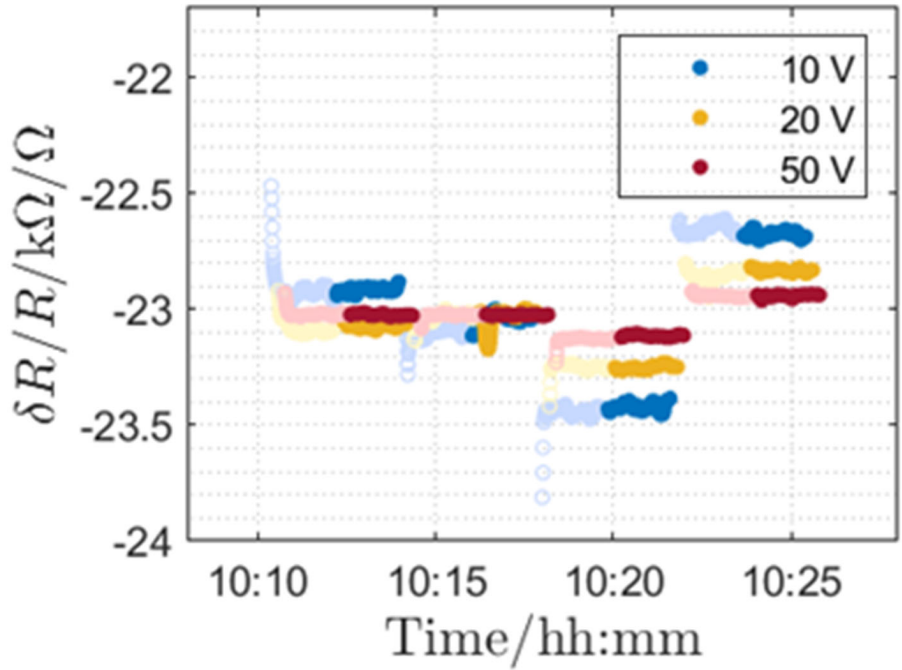

Fig. 2. Several data acquisition runs showing the stability at 10 V, 20 V, and 50 V for the 98.358 MΩ artifact resistor network. Light colors indicate data collected during 120 s settling time, dark colors indicate data used for determination of the resistance.

For the second case, artifacts resistors were used to build and emulate the network in Fig. 1 (e). When measured at 50 V (corresponding to approximately 50 nA) with a teraohmmeter, the effective resistance was approximately 1.02 GΩ (yielding an error of less than 1%). A dual source bridge was also used to measure a second artifact resistor network, trimmed closer to 1.01 GΩ, which confirmed the teraohmmeter measurements.

Since these networks can be measured reliably with the techniques mentioned, one may proceed with fabricating QHARS devices that can output highly accurate values of quantized resistance. Ultimately, this work sets the foundation for achieving a means of shortening traceability chains for high resistance calibrations.

REFERENCES

1. T. Oe, S. Gorwadkar, T. Itatani and N. Kaneko, "Development of 1 MΩ Quantum Hall Array Resistance Standards," *IEEE Transactions on Instrumentation and Measurement*, 66, 6, 1475-1481, 2017.
2. D. S. Scaletta, S. M. Mhatre, N. T. M. Tran, C. H. Yang, H. M. Hill, Y. Yang, L. Meng, A. R. Panna, S. U. Payagala, R. E. Elmquist, D. G. Jarrett, D. B. Newell, and A. F. Rigosi, "Optimization of graphene-based quantum Hall arrays for recursive star–mesh transformations", *Appl. Phys. Lett.* 123, 153504, 2023.
3. A. F. Rigosi, A. L. Levy, M. R. Snure, N. R. Glavin, *J. Phys. Mater.* 4, 032003, 2021.
4. H. M. Hill *et al.*, *Phys. Rev. B*, 98, 165109, 2018.
5. M. Kruskopf *et al.*, *IEEE Trans. Electron Dev.*, 68, 3672-3677, 2021.
6. X. Wang, E. Khatami, F. Fei, J. Wyrick, P. Namboodiri, R. Kashid *et al.*, *Nat. Commun.*, 13, 6824, 2022.
7. U. Wurstbauer, D. Majumder, S. S. Mandal, I. Dujovne, T. D. Rhone *et al.*, *Phys. Rev. Lett.*, 107, 066804, 2011.
8. D. G. Jarrett, C. C. Yeh, S. U. Payagala, A. R. Panna, Y. Yang, L. Meng, S. M. Mhatre, N. T. M. Tran, H. M. Hill, D. Saha, R. E. Elmquist, D. B. Newell, A. F. Rigosi, "Graphene-based star-mesh resistance networks", *IEEE Transactions on Instrumentation and Measurement*, 72, 1502710, 1-10, 2023.
9. A.E. Kennelly, "Equivalence of triangles and stars in conducting networks", *Electrical World and Engineer*, 34, 413–414 (1899).